\documentclass[aps,preprint]{revtex4}
\def\beq{\begin{equation}} \def\eeq{\end{equation}}
\usepackage{graphics,epsfig,amssymb}

\begin{document}

\title{Alpha-like quartet condensation and isovector pairing correlations in N=Z nuclei}

\author{N. Sandulescu, D. Negrea}
\affiliation{
National Institute of Physics and Nuclear Engineering, P.O. Box MG-6, 76900 Bucharest-Magurele, Romania}
\author{ J. Dukelsky}
\affiliation{
Instituto de Estructura de la Materia, CSIC, Serrano 123, 28006 Madrid, Spain}
\author{ C. W. Johnson}
\affiliation{Department of Physics, San Diego State University,
 5500 Campanile Drive, San Diego, CA 92182-1233}

\begin{abstract}

We propose a simple quartet condensation model (QCM) which describes with
very high accuracy the isovector pairing correlations in self-conjugate
nuclei. The quartets have an alpha-like structure and are formed by
collective isovector pairs. The accuracy of the QCM is tested
for N=Z nuclei for which exact shell model diagonalizations can
be performed. The calculations are done with two isovector pairing forces,
one extracted from standard shell model interactions and the other of
seniority type, acting, respectively, upon spherical and axially-deformed
single-particle states. It is shown that for all calculated nuclei the
QCM gives very accurate values for the pairing correlations energies,
with errors which do not exceed 1$\%$. These results show clearly that the
correlations induced by the isovector pairing in self-conjugate nuclei
are of quartet type and also indicate that  QCM is the proper tool to
calculate the isovector proton-neutron correlations in mean field
pairing models.

\end{abstract}

\maketitle

\section{introduction}

Pairing and quantum condensation are outstanding  phenomena in many domains of physics.
The best known pairing phenomenon  is the superconductivity of electrons in metals,
described by Bardeen-Cooper-Schrieffer (BCS) theory \cite{bcs}. For more than
50 years BCS-type approaches
have also been applied to describe superfluid properties of atomic nuclei
and neutron stars. However, compared to electronic systems, nuclear
systems present special features because they involve two kinds of fermions,
neutrons and  protons.
Therefore  in nuclear systems one can have not only Cooper pairs of like fermions,
such as neutron-neutron (nn) and proton-proton (pp) pairs , but also two types
of proton-neutron (pn)
pairs, i.e.,  isovector (isospin T=1) and isoscalar (isospin T=0) pairs.
All these pairs should be equally considered in the nuclear systems with the
same number of neutrons and protons, which is a difficult task for nuclear
microscopical models.

A common approach to treat pairing in N=Z nuclei, where N and
Z are the numbers of neutrons and protons, is the generalized
Hartree-Fock-Bogoliubov approximation
(HFB)  \cite{goodman}. In this approach all types of Cooper pairs are treated in a unified
manner but with a trial wave function which does not conserve exactly the
particle number and the isospin of the nucleus. Here we shall investigate
another approach based not on Cooper pairs but on 4-body clusters composed
of two neutrons and two protons coupled to the isospin T=0 and to
the angular momentum J=0. This 4-body structure is commonly called an
alpha-like quartet. The existence of alpha-like quartets in nuclei is
a long standing issue \cite{oerzen}. Quartet structures are also under current
investigations in other physical systems, such as spin-3/2 cold fermionic atoms,
two band electronic systems or bi-layered  systems with particles
and holes \cite{wu,moskalenko}.

Various studies have  raised the question if a condensate of alpha-like
quartets could exist in the ground state of N=Z nuclei
\cite{flowers,eichler,gambhir,dobes,hasegawa, zelevinsky}.
One of the first microscopic models of  quartet condensation in nuclei was
proposed by Flowers et al \cite{flowers} and it was based on a BCS-type
function written in terms of quartets. Recently a similar calculation
scheme  was proposed by including in the BCS function both pairs and
quartets \cite{zelevinsky}. A theory of quartet condensation based on
a BCS-type function  has the advantage of simplicity but its applicability
to real nuclei is hindered by the fact that it does not conserve exactly
the particle number, which in the case of quarteting is uncertain in groups
of 4 particles at a time. This limitation was discussed extensively in
Ref. \cite{dobes} for the particular case of a degenerate shell.
A quartet condensation approach which conserves the number of particles
has been proposed in Ref \cite{eichler}. In this approach quartets
are constructed for each single particle level, which makes the
calculations cumbersome. A more general calculation scheme, based
on a simplified version of the quartet model \cite{arima} and which
works accurately for systems with two quartets outside a closed
core, was proposed in Ref. \cite{hasegawa}. Quartet condensation was
also  analyzed in Ref. \cite{gambhir}  by considering phenomenological
 bosons.

The scope of the present paper is to introduce a simple quartet condensation
model formulated in terms of collective Cooper pairs which is able
to describe with a very good accuracy the isovector pairing correlations
in nuclei.

The isovector pairing correlations are described by the Hamiltonian
\beq
\hat{H}= \sum_i \varepsilon_i (N^{\nu}_i+N^{\pi}_i)
- \sum_{i,j,\tau} V_{ij} P^+_{i,\tau} P_{j,\tau},
\eeq
where the first term is the single-particle part and the second
is the most general isovector interaction. The isovector interaction is
invariant under rotations in isospace and it is expressed in terms
of the isovector pair operators $P^+_{i,1}=\nu^+_i \nu^+_{\bar{i}}$, 
$P^+_{i,-1}=\pi^+_i \pi^+_{\bar{i}}$ and 
$P^+_{i,0}=(\nu^+_i \pi^+_{\bar{i}} + \pi^+_i \nu^+_{\bar{i}})/\sqrt{2}$; the
operators $\nu^+_i$ and $\pi^+_i$ create, respectively, a neutron and a proton in 
the state $i$ while $\bar{i}$ denotes the time conjugate of the state $i$.
When all the matrix elements of the interaction are considered of equal strength, 
the Hamiltonian (1) has SO(5) symmetry and its exact solutions, both for a degenerate 
and a non-degenerate single-particle spectrum, have been discussed extensively in 
the literature (e.g., see \cite{richardson,engel,link,dukelsky}).

 The ground state of the Hamiltonian (1) is described here in terms of
alpha-like quartets. First, we introduce the non-collective quartet operators $A^+_{ij}$
obtained by coupling two isovector pairs to the total isospin T=0, i.e.,
\beq
A^+_{ij} = [P^+_i P^+_j]^{T=0} = \frac{1}{\sqrt 3}(P^+_{i,1} P^+_{j,-1}+P^+_{i,-1} 
P^+_{j,1}
           -P^+_{i,0} P^+_{j,0})
\eeq
With these operators we construct a collective quartet operator
\beq
A^+ = \sum_{i,j} x_{ij} A^+_{ij}
\eeq
where the summation is over the single-particle states included in the calculations.

Finally, with the collective quartet operator we construct the quartet condensate
\beq
| \Psi \rangle =(A^+)^{n_q}|0 \rangle
\eeq
where $n_q$ is the number of quartets. By construction, this wave function
has N=Z=2n$_q$ and a well-defined total isospin T=0. In addition, if the
Hamiltonian (1) has spherical (axial) symmetry, the quartet condensate (4) 
has also J=0 (J$_z$=0), where J and J$_z$ are the total angular momentum
and its projection on the symmetry axis.

We would like to stress that in this study the word  {\it condensate} means
that the function (4)  is obtained by applying the same quartet operator many times .
Since the quartet operator is a composite fermionic operator,
the condensate wave function (4) is  not a bosonic condensate of alpha  particles.
In fact, the quartets considered in the present model are 4-body structures
correlated in angular momentum and isospin space rather than tight alpha
clusters correlated in coordinate space.

The calculations with the function (4) can be greatly simplified if we assume that
the mixing amplitudes $x_{ij}$ which define the collective quartet operator (3)
have a separable form, i.e., $x_{i,j}=x_{i} x_{j}$. In this case the collective
quartet operator can be written as
\beq
A^+= 2 \Gamma^+_1 \Gamma^+_{-1} - (\Gamma^+_0)^2
\eeq
where $\Gamma^+_{\tau}= \sum_i x_i P^+_{i\tau}$ are the collective Cooper pair operators
corresponding to the nn, pp and np pairs. Due to the isospin invariance, all the
collective pairs have the same mixing amplitudes $x_i$. With the operator (5) 
the state (4) can be written in the following form
\begin{eqnarray}
| \Psi \rangle &=& (2 \Gamma^+_1 \Gamma^+_{-1} - \Gamma^{+2}_0 )^{n_q} |0 \rangle \nonumber \\
         &=& \sum_k
\left(\begin{array}{l} n_q \\ k \end{array}\right)
(-1)^{n_q-k} 2^k (\Gamma^+_1 \Gamma^+_{-1})^k \Gamma^{+2(n_q-k)}_0 |0\rangle
\label{psi}
\end{eqnarray}
One can notice that in the expansion above there are two terms which
correspond to two particle-number-projected BCS (PBCS) wave functions
\beq
|PBCS0 \rangle = \Gamma^{+2 n_q}_0 |0 \rangle
\eeq
\beq
| PBCS1 \rangle = (\Gamma^+_1 \Gamma^+_{-1})^{n_q} |0 \rangle
\eeq
The function (7) is a condensate of proton-neutron
pairs while the function (8) is a product of a condensate
of neutron-neutron pairs with a condensate of proton-proton  pairs.
Both PBCS functions conserve the number of particles and the
projection of the total isospin on z-axis, but they do not have
a well-defined total isospin. As seen from the structure of the
quartet condensate (6), in order to restore
the isospin symmetry one needs to take a combination of all PBCS functions
with the number of pairs compatible with the binomial expansion.

The quartet condensate (6) is defined by the mixing amplitudes $x_i$. They are
found from the minimization of the average of the Hamiltonian,
$ \langle \Psi | H | \Psi \rangle$, and from the normalization
condition $ \langle \Psi | \Psi \rangle =1$. The average of the Hamiltonian
and the norm are calculated using recurrence relations. This method
is based on the relations satisfied by the matrix elements of the pairing interaction
between states with an arbitrary number of nn, pp and np collective pairs
defined by
\beq
| n_1 n_2 n_3 \rangle = \Gamma^{+n_1}_1 \Gamma^{+n_2}_{-1} \Gamma^{+n_3}_0 |0 \rangle
\eeq
As an example, we give below the recurrence relations satisfied
by the matrix elements of the operator $P^+_{i1} P_{j1}$:
\begin{eqnarray*}
&&\langle n_1 n_2 n_3|P^+_{i,1}P_{j,1}|m_1 m_2 m_3\rangle = x^2_i x^2_j \{
n_1 m_1 n_11 m_11 \langle n_12 n_2 n_3 |P^+_{j,1}P_{i,1}|m_12 m_2 m_3 \rangle \\
&+& \frac{1}{4} n_3 m_3 n_31 m_31 \langle n_1 n_2 n_32|P^+_{j,-1}P_{i,-1} |m_1 m_2 m_32 \rangle \\
&+& n_1 m_1 m_3 ( n_3  \langle n_11 n_2 n_31 |P^+_{j,0}P_{i,0}|m_11 m_2 m_31 \rangle 
+ n_11 \langle n_12 n_2 n_3 |P^+_{j,0}P_{i,1}|m_11 m_2 m_31 \rangle ) \\
&+& \frac{1}{2} n_3 m_1 n_31 ( m_3 \langle n_1 n_2 n_32|P^+_{j,0}P_{i,-1}|m_11 m_2 m_31 
\rangle 
+ m_11 \langle n_1 n_2 n_32|P^+_{j,1}P_{i,-1}|m_12 m_2 m_3 \rangle ) \\
&+& \delta_{ij} [ n_1 n_3 m_1 m_3 (\langle n_11 n_2 n_31|m_11 m_2 m_31\rangle
 - \frac{1}{2} \langle n_11 n_2 n_31|N^{\nu}_i+N^{\pi}_i|m_11 m_2 m_31 \rangle) \\
&+& n_1 m_1 n_11 m_11 (\langle n_12 n_2 n_3|m_12 m_2 m_3\rangle 
 - \langle n_12 n_2 n_3|N^{\nu}_i|m_12 m_2 m_3\rangle) \\
&+& \frac{1}{4} n_2 m_3 n_31 m_31(\langle n_1 n_2 n_32|m_1 m_2 m_32\rangle
 - \langle n_1 n_2 n_32|N^{\pi}_i|m_1 m_2 m_32\rangle ) \\
&-& n_1 m_1 (m_3 n_11 \langle n_12 n_2 n_3|T_{i,-1}|m_11 m_2 m_31\rangle 
- n_3 m_11 \langle m_12 m_2 m_3|T_{i,-1}|n_11 n_2 n_31\rangle ) \\
&+& \frac{1}{2} n_3 m_3 (n_31 m_1 \langle n_1 n_2 n_32|T_{i,1}|m_11 m_2 m_31\rangle 
+ n_1 m_31 \langle m_1 m_2 m_32|T_{i,1}|n_11 n_2 n_31\rangle ) ] \} \\
&+& m_1 x_j \langle m_11 m_2 m_3|P_{i,1}|n_1 n_2 n_3\rangle 
- x_i x^2_j [ n_1 m_1 m_3\langle m_11 m_2 m_31|P_{j,0}|n_11 n_2 n_3\rangle \\
&+& n_1 m_1 m_11\langle m_12 m_2 m_3|P_{j,1}|n_11 n_2 n_3\rangle 
+ \frac{1}{2}n_1 m_3 m_31 \langle m_1 m_2 m_32|P_{j,-1}|n_11 n_2 n_3\rangle ]
\end{eqnarray*}
In the expressions above $T_{i,\tau}$ are the isospin operators and $n_ik$ denotes
$n_i-k$(e.g., $n_i1=n_i-1$) From this example it can be seen
that by the recurrence relations are generated not only the matrix elements for
the pairing interaction but also for the operators $P^+_{i,1}P_{j,-1}$,
$P^+_{i,0}P_{j,1}$, $P^+_{i,0}P_{j,-1}$. The matrix elements of all these operators
depend on  each other through the recurrence relations satisfied by each of them. 
More details about the calculation scheme will be given in a 
forthcoming publication.

It is worth mentioning that the state (4) with the quartet (5) has a similar form
with the eigenstate of zero seniority employed in the generalized-seniority model (GSM)
of definite isospin \cite{talmi}. However, since the calculation scheme used here 
is valid for any isovector pairing interaction while GSM is valid only for the
interactions which satisfy specific commutation relations with the pair operators,
the present quartet model and the GSM are different, with the former
being more general.

The quartet condensation model (QCM) introduced above is applied here with
isovector interactions and single-particle spectra commonly used in shell
model and mean field calculations (for a preliminary study
with schematic spectra see Ref. \cite{sandulescu_qcm})  
 To check the accuracy of QCM we have chosen three sets of N=Z nuclei for which
 exact shell model calculations can be performed. The three sets of nuclei,
shown in Table I between horizontal lines,  have the valence nucleons moving outside
the double-magic cores $^{16}$O, $^{40}$Ca and $^{100}$Sn. These cores are considered
as inert in the present calculations and the valence nucleons are described with the
Hamiltonian (1). First we have applied the QCM for an isovector pairing force
extracted from the (T=1,J=0) part of standard shell model interactions acting on
spherical single-particle states. More precisely, for the
three sets of nuclei shown in Table I we have used, respectively, the following sources
for the isovector pairing matrix elements (all other matrix elements were set to zero):
1) a universal $sd$-shell interaction (USDB) \cite{usd} and
the energies (in MeV): $\varepsilon_{1d_{5/2}}$=-3.926,
$\varepsilon_{2s_{1/2}}$=-3.208,$\varepsilon_{1d_{3/2}}$=2.112;
2) a monopole-modified Kuo-Brown interaction (KB3G) \cite{kb3} and the energies
$\varepsilon_{1f7/2}=0.0$, $\varepsilon_{2p3/2}=2.0$, $\varepsilon_{2p1/2}=4.0$,
$\varepsilon_{1f5/2}=6.5$;
3) the effective G-matrix interaction of Ref. \cite{bonnA} and the
energies $\varepsilon_{2d_{5/2}}$=0.0, $\varepsilon_{1g_{7/2}}$=0.2,
$\varepsilon_{2d_{3/2}}$=1.5, $\varepsilon_{3s_{1/2}}$=2.8.
In order to be able to perform exact shell model calculations in the major
shells N=28-50 and N=50-82, the single-particle states $1g_{9/2}$ and
$1h_{11/2}$ were not introduced in the calculations. These limitations
do not apply to the quartet model calculations, which can be done for
larger N=Z systems than can be presently calculated by the shell model.

\begin{table}[hbt]
\caption{ Correlation energies calculated with isovector pairing forces
extracted from standard shell model interactions and with spherical 
single-particle states. The results correspond to exact shell model
diagonalisations (SM), quartet condensation model (QCM), 
and the two PBCS approximations of Eqs.(7,8).
Numbers in the brackets are the errors relative to the exact shell model results. }
\begin{center}
\begin{tabular}{|c|c|c|c|c|}
\hline
\hline
   &    SM & QCM & PBCS1 & PBCS0  \\
\hline
\hline
$^{20}$Ne  &  9.173    & 9.170  (0.033\%)  & 8.385 (8.590\%)  & 7.413 (19.187\%) \\
$^{24}$Mg  &  14.460   & 14.436 (0.166\%)  & 13.250 (8.368\%) & 11.801 (18.389\%) \\
$^{28}$Si  &  15.787   & 15.728 (0.374\%)  & 14.531 (7.956\%) & 13.102 (17.008\%) \\
$^{32}$S   &  15.844   & 15.795 (0.309\%)  & 14.908 (5.908\%) & 13.881 (12.389\%) \\
\hline
$^{44}$Ti  &  5.973    & 5.964 (0.151\%)   & 5.487 (8.134\%)  & 4.912 (17.763\%) \\
$^{48}$Cr  &  9.593    & 9.569 (0.250\%)   & 8.799 (8.277\%)  & 7.885 (17.805\%) \\
$^{52}$Fe  &  10.768   & 10.710 (0.539\%)  & 9.815 (8.850\%)  & 8.585 (20.273\%)  \\
\hline
$^{104}$Te &  3.831    & 3.829 (0.052\%) & 3.607 (5.847\%) & 3.356 (12.399\%) \\
$^{108}$Xe &  6.752    & 6.696 (0.829\%) & 6.311 (6.531\%) & 5.877 (12.959\%) \\
$^{112}$Ba &  8.680    & 8.593 (1.002\%) & 8.101 (6.670\%) & 13.064 (13.064\%) \\

\hline
\hline
\end{tabular}
\end{center}
\end{table}

The results for the correlations energies, defined as $E_{corr}=E_0-E$,
where $E$ is the energy of the ground state and $E_0$ is the
energy calculated without taking into account the isovector
pairing interaction, are shown in Table I.
In the second column are shown the exact shell model results, in the next
column the results of QCM while in the last two column are given the results
of the two PBCS approximations defined by Eqs.(7,8). In the brackets
are given the errors relative to the exact shell model results.
Concerning the PBCS approximations, it can be seen that the lowest energy
is obtained for the state PBCS1 and not for the proton-neutron condensate PBCS0.
The latter gives the lowest energy for N=Z=odd systems \cite{sandulescu_errea}.
It can be also noticed that the errors  corresponding to the PBCS approximations
are rather large, much larger than for pairing between like particles
\cite{sandulescu_bertsch}.

The most remarkable result seen in Table I is that QCM
gives very small errors, of below $1\%$, for
all the calculated isotopes, including the ones with 3 and 4
quartets. It should be mentioned also that QCM  calculations
are also very fast (few CPU  minutes on an ordinary laptop) 
and can be applied for nuclei with many proton-neutron pais,
which cannot be calculated with the present SM codes.

An interesting issue is the relation between the QCM and the
BCS calculations, called below PBCS(N,T), in which both the
particle number and the total isospin are restored using
projection techniques. To address this issue we consider
here the PBCS(N,T) result for $^{52}$Fe shown in Table 3
of Ref. \cite{chen}. The PBCS(N,T) calculations are done with
an isovector pairing force of constant strength, with a value
equal to g=-24/A, where A is the mass of the nucleus, and with
spherical single-particle states (for details, see Ref. \cite{chen})
The correlation energy obtained with the PBCS(N,T) approximation
is 7.63 MeV, which should be compared to the exact value, equal
to 8.29 MeV. With the same input the correlation energy obtained
with the QCM is equal to 8.25 MeV. This comparison shows that the
QCM is much more accurate than PBCS(N,T) and indicates also that
QCM describes additional, quartet-type correlations which cannot
be  obtained in the standard BCS-type models.

\begin{table}[hbt]
\caption{ Correlation energies calculated with an isovector pairing
force of seniority type and with axially-deformed single-particle
states. The notations are the same as in Table 1.}
\begin{center}
\begin{tabular}{|c|c|c|c|c|}
\hline
\hline
   &    SM & QCM & PBCS1 & PBCS0  \\
\hline
\hline
$^{20}$Ne  &  6.55     & 6.539  (0.168\%)  & 5.752  (12.183\%) &  4.781 (27.008\%) \\
$^{24}$Mg  &  8.423    & 8.388  (0.415\%)  & 7.668  ( 8.963\%) & 6.829  (18.924\%) \\
$^{28}$Si  &  9.661    & 9.634  (0.279\%)  & 9.051  ( 6.314\%) & 8.384  (13.218\%) \\
$^{32}$S   &  10.263   & 10.251 (0.117\%)  & 9.854  ( 3.985\%) & 9.372  (18.682\%) \\
\hline
$^{44}$Ti  &  3.147    & 3.142  (0.159\%)  & 2.750  (12.615\%) & 2.259  (28.217\%) \\
$^{48}$Cr  &  4.248    & 4.227  (0.494\%)  & 3.854  ( 9.275\%) & 3.423  (19.421\%) \\
$^{52}$Fe  &  5.453    & 5.426  (0.495\%)  & 5.033  ( 7.702\%) & 4.582  (15.973\%)  \\
\hline
$^{104}$Te & 1.084     & 1.082  (0.184\%)  & 0.964 (11.070\%) &  0.832 (23.247\%) \\
$^{108}$Xe & 1.870     & 1.863  (0.374\%)  & 1.697 (9.264\%) & 1.514  (19.037\%) \\
$^{112}$Ba &  2.704    & 2.688  (0.592\%)  & 2.532  (6.361\%) & 2.184 (19.230\%) \\
\hline
\hline
\end{tabular}
\end{center}
\end{table}

In N=Z nuclei there are  other important degrees of freedom
which compete with the isovector interaction. This can be easily
seen from  the small overlap between the exact shell model wave 
functions calculated with the isovector pairing force and with 
the full two-body interaction. For example, in the case of $^{48}$Cr
this overlap is equal to 0.614, which is a very small value when
compared to similar overlaps calculated for spherical nuclei with
like-particle pairing \cite{sandulescu_gsen}. Among the most
important degrees of freedom which compete with the pairing
in N=Z nuclei are the quadrupole ones. It has been shown that
a simplified model Hamiltonian which includes only the isovector
pairing interaction and a quadrupole-quadrupole force is able
to give a realistic description of the essential features 
of N=Z nuclei \cite{langanke,pittel} In the mean field version
of this model the quadrupole degrees of freedom are commonly 
included in a deformed mean field and the isovector pairing 
is treated in the single-particle basis corresponding to it
(e.g., see \cite{bentley})
In what follows we use this framework to analyze the quartet 
correlations in deformed N=Z nuclei. As an illustration we consider 
the same nuclei shown in Table 1. The QCM is applied for 
an isovector pairing interaction acting on the single-particle 
spectrum corresponding to an axially deformed mean field. 
The mean field is generated self-consistently by Hartree-Fock 
calculations performed with the Skyrme force SLy4 \cite{sly4} and 
using the code ev8 \cite{ev8}.
For the isovector pairing force we take a seniority type interaction
with the strength g=-24/A \cite{chen}. To keep the analogy with the 
calculations done above for a Hamiltonian with spherical symmetry, 
the Coulomb interaction is neglected in the mean field and the pairing
is applied for the Hartree-Fock (HF) single-particle 
states above the  cores $^{16}$O, $^{40}$Ca 
and $^{100}$Sn. More precisely, to be able to perform exact shell 
model diagonalizations, for the three sets of nuclei shown in Table 1 
we consider, respectively, the lowest 7, 9 and 10 deformed HF 
single-particle states above the double-magic cores. 
The results, given in Table 2, show that  QCM gives  
very accurate results for a deformed mean field.   
Although it is not shown, when we vary the interaction strength,
from the weak to the strong coupling regime,
we continue to get similar very good accuray.

In conclusion, we have shown that the isovector pairing correlations in
N=Z nuclei are described with a very high accuracy by a condensate of 
alpha-like quartets formed by collective pairs. Because its accuracy
and simplicity,  the QCM appears to be the appropriate tool for 
describing the isovector  pairing correlations in nuclei. 

Finally we would like to mention that due to the general structure
of the state (9) employed in the recurrence relations, which is defined
for an arbitray number of pairs, the QCM can be also extended to treat
nuclei with a different number of protons and neutrons in the same 
open shell. The study of quartet correlations in such nuclei is the
scope of a future investigation.

\vskip 0.2cm
\noindent
{\bf Acknowledgements}
\vskip 0.1cm
\noindent
N.S thanks Stuart Pittel for valuable discussions and Denis Lacroix for 
the help in running the code ev8.
This work was supported by the Romanian Ministry of Education 
and Research through the grant Idei nr 57, by the Spanish Ministry 
for Science and Innovation Project No. FIS2009-07277 and by 
the U.S. Department of Energy through the grant DE-FG02-96ER40985. 
C.W.J and N.S thank the Institute for Nuclear Theory at the 
University of Washington, where their collaboration have 
been initiated.

\end{document}